%% file: Na20.tex
\documentclass[aps,prc,twocolumn,superscriptaddress,amsmath,amssymb,showpacs,nofootinbib]{revtex4}
\usepackage{graphicx,color}
\usepackage{ulem}
\usepackage{multirow}

\usepackage{enumerate}
\usepackage{natbib}

\newcommand\T{\rule{0pt}{3.1ex}}
\newcommand\B{\rule[-1.7ex]{0pt}{0pt}}

\begin{document}

\title{High-statistics measurement of the $\boldsymbol{\beta}$-delayed $\boldsymbol{\alpha}$ spectrum of $^{\boldsymbol{20}}$Na}

\author{K.~L.~Laursen}
\email[Corresponding author: ]{klj06@phys.au.dk}
\affiliation{Department of Physics and Astronomy, Aarhus University, DK-8000 Aarhus C, Denmark}

\author{O.~S.~Kirsebom}
\affiliation{Department of Physics and Astronomy, Aarhus University, DK-8000 Aarhus C, Denmark}
\affiliation{TRIUMF, Vancouver, British Columbia, V6T 2A3, Canada}

\author{H.~O.~U.~Fynbo}
\affiliation{Department of Physics and Astronomy, Aarhus University, DK-8000 Aarhus C, Denmark}

\author{A.~Jokinen}
\affiliation{Department of Physics, University of Jyv{\"a}skyl{\"a}, FIN-40351 Jyv{\"a}skyl{\"a}, Finland}

\author{M.~Madurga}
\altaffiliation[Present address: ]{Physics Division, Oak Ridge National Laboratory, Oak Ridge, TN 37831, USA.}
\affiliation{Instituto de Estructura de la Materia, CSIC, Serrano 113 bis, E-28006 Madrid, Spain}

\author{K.~Riisager}
\affiliation{Department of Physics and Astronomy, Aarhus University, DK-8000 Aarhus C, Denmark}

\author{A.~Saastamoinen} 
\altaffiliation[Present address: ]{Cyclotron Institute, Texas A\&M University, College Station, TX 77843-3366, USA}
\affiliation{Department of Physics, University of Jyv{\"a}skyl{\"a}, FIN-40351 Jyv{\"a}skyl{\"a}, Finland}

\author{O.~Tengblad}
\affiliation{Instituto de Estructura de la Materia, CSIC, Serrano 113 bis, E-28006 Madrid, Spain}

\author{J.~{\"A}yst{\"o}}
\affiliation{Department of Physics, University of Jyv{\"a}skyl{\"a}, FIN-40351 Jyv{\"a}skyl{\"a}, Finland}
\affiliation{Helsinki Institute of Physics, FI-00014 University of Helsinki, Finland}	


\begin{abstract}
  A measurement of the $^{20}$Na $\beta$-delayed alpha spectrum with a high-granularity set-up has allowed the decay scheme to be
  revised on several points. Three new transitions of low intensity
  are found at low $\alpha$-particle energy. An $R$-matrix fit of the
  complete spectrum gives an improved description of the decay and
  indicates feeding to the broad $2^+$ $\alpha$-cluster state close to
  9~MeV.
\end{abstract}




\maketitle



\section{Introduction}

The $J^{\pi}=2^+$, $T=1$ ground state of $^{20}$Na decays to $^{20}$Ne
by positron emission with a half-life of 447.9(23)~ms and an available decay energy of
$Q_{\textrm{EC}} = 13\, 892.5(11)$~keV~\cite{Wang2012}. Excited states in
$^{20}$Ne populated in the $\beta$ decay of $^{20}$Na, with an
excitation energy above $4\, 729.84\textrm{~keV}$, may break up to
$\alpha+{}^{16}\textrm{O}$. The $\alpha$ particles originating from
such transitions are referred to as $\beta$-delayed $\alpha$ particles
and have been the subject of numerous experimental investigations.

Through angular correlation studies, see e.g. Refs.~\cite{macfarlane71,
  clifford89}, the ratio of the Fermi (vector) and the Gamow-Teller
(axial-vector) component has been determined
for a number of allowed transitions including the $J^{\pi}=2^+$,
$T=1$ isobaric analog state (IAS) at $10\, 273$~keV. 
For several transitions to $J^{\pi}=2^+$, $T=0$ states in $^{20}$Ne,
limits have been set on the Fermi contribution and from this the $^{20}$Ne isospin mixing has also been established.

Through measurements of $ft$ values to individual levels in $^{20}$Ne and comparison to the
$ft$ values in the mirror decay, $^{20}\textrm{F}\rightarrow
{}^{20}\textrm{Ne}^{\ast}+\beta^{+}+\nu_e$, the energy-dependence of
the mirror asymmetry has been studied, yielding information on nuclear
structure, meson exchange currents, and second-class currents, see e.g. Refs.~\cite{torgerson73, clifford89}.

The data presented here were originally intended as calibration data for a
precision measurement of the $\beta$ decay of $^8$B~\cite{kirsebom11_8B}, but the
quality and high statistics of the data reveals new features of the
decay scheme of $^{20}$Na. New transitions appear at low $\alpha$
particle energies and clear signatures of
interference leads to a reinterpretation of the decay scheme at higher $\alpha$-particle energies. In particular, we shall present evidence for the
population of a broad resonance in $^{20}$Ne, indicative of a
$\alpha+{}^{16}$O cluster structure. It can be studied in a very
clean way through $\beta$ decay since the initial state and the
transition operator are well understood so that the final state is the
only unknown, see Ref.~\cite{hyldegaard10_prc} for a similar study in the
$A=12$ system.

The broad (cluster) states can play a role in the
$^{16}$O($\alpha,\gamma$)$^{20}$Ne reaction, which proceeds at a slow rate during astrophysical He burning due to the absence of natural-parity resonances inside the Gamow window. Recently Hager \textit{et
  al.}  \cite{Hager} measured the total $S$-factor at $E_\text{c.m.} =
2.26\, \text{MeV}$ and determined the contributions from direct
capture transitions to the ground and first-excited state. Costantini
\textit{et al.} \cite{Costantini} determined $S$-factor values for a
range of energies and used an $R$-matrix fit to extrapolate the $S$-factor
down to the energies relevant for astrophysics, but they left out the
broad $0^+$ and $2^+$ resonances of the fourth $0^+$ rotational band
of $^{20}$Ne.  Our results may allow to judge more precisely the
effect on the $S$-factor of the low-energy tail of the $2^+_4$ resonance.

\section{Experiment}

\subsection{Beam production}

The radioactive $^{20}$Na beam was produced at the Accelerator
Laboratory of University of Jyv\"askyl\"a through the
$^{24}\textrm{Mg}(p,n\alpha){}^{20}\textrm{Na}$ reaction.  The proton
primary beam had an energy of 40~MeV and a typical intensity of
8\,--\,10~$\mu$A, and the target was self-supporting 4.3~mg/cm$^2$ natural
Mg.  The activity was extracted with the Ion Guide Isotope Separation
On-Line (IGISOL) technique~\cite{aysto01}, using a light-ion fusion
ion guide~\cite{huikari04} and the mass separated $A/q=20$ beam was
implanted in a carbon foil of thickness $26\pm 2$~$\mu$g/cm$^2$. The
acceleration voltage was 20~kV resulting in the ions, on average,
being implanted at a depth of 7.4~$\mu$g/cm$^2$.  The average
$^{20}$Na implantation rate was $2.2\times 10^4$ ions per second.

Data taking was split in two runs, the first lasting 5~hours, the
second 7~hours. In between, the setup was used for measuring the
decays of $^{23}$Al (8~hours) and $^8$B (72 hours). At no point during
this interval was the vacuum broken. The $^{8}$B and $^{23}$Al
measurements are discussed
elsewhere~\cite{kirsebom11_8B,kirsebom11_al23}. The two $^{20}$Na runs
give consistent results.

\subsection{Detection system}

\begin{figure}
  \centering
        	\includegraphics[width=1.0\columnwidth]{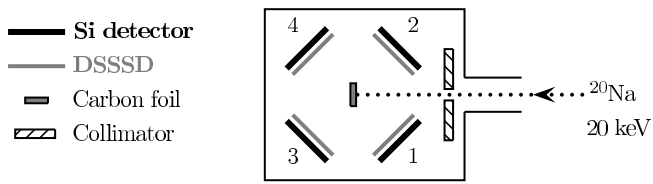}
  \caption{Schematic and simplified illustration of the experimental setup (top view).}
  \label{fig : setup}
\end{figure}

The set-up consisted of four 60 $\mu$m thick, double sided silicon strip
detectors (DSSSD)~\cite{tengblad04} backed by 1.5 mm thick,
unsegmented silicon detectors used to veto against $\beta$ particles.
The detectors were placed 5~cm from the carbon foil in a rectangular
configuration with the carbon foil perpendicular to the beam, see
Fig.~\ref{fig : setup}. The total solid-angle coverage was 30\% and
the angular resolution 3~degrees.
An important feature of the DSSSDs
is the very thin dead layer of only 100~nm
(over 95\,-\,96\% of the active surface) which facilitates the detection of
low-energy ions~\cite{tengblad04}. 
The energy resolution of the DSSSDs was 25 keV (FWHM).

The trigger thresholds of the data acquisition
system were set as low as possible above the
noise and were studied with an $^{241}$Am
source in air at atmospheric pressure~\cite{kirsebom11_8B}. 
The trigger efficiency was found to rise gently as a function of energy, increasing from 0\% to 100\% within an interval of roughly 100~keV. The trigger thresholds, defined as the energy at which the efficiency reaches 50\%, ranged from 160 to 240~keV, depending on the electronics channel.
Low-energy particles below the trigger threshold could be detected if
in coincidence with a higher-energy particle. Low-energy cutoffs in each ADC channel 
ranged from 60 to 230~keV.

\subsection{Energy calibration}

The two most intense $\beta$-delayed $\alpha$ lines of $^{20}$Na at
2153.2(10)~keV and 4433.8(15)~keV were used for the energy calibration. 
Their energies were deduced from the excitation energies,
7421.9(12)~keV and 10273.2(19)~keV, of the corresponding states in
$^{20}$Ne and the $\alpha+{}^{16}\textrm{O}$ threshold energy of
4729.84(1)~keV given in Ref.~\cite{tunl_A20}.
SRIM stopping power tables~\cite{SRIM} were used to correct
for the energy loss of the $\alpha$ particles and the $^{16}$O
ions in the carbon foil and the detector dead layer, taking
into account the variation in effective thickness with angle, but
assuming a fixed implantation depth in the foil.
Corrections were also made
for the non-ionizing energy loss in the active volume of the detector
which does not contribute to the observed signal. 
The calibration includes a 5~keV quadratic component due to a
non-linearity in the electronical response (deduced from measurements
with a precision pulse generator) and a 20~keV quadratic component
due to the changing pulse shape of physical particles at low
energies. See Ref.~\cite{kirsebom11_8B} for further details.

\input{sec3_lowenergy}

\section{Summary and conclusion}
\begin{figure*}
\includegraphics[width=1.0\textwidth]{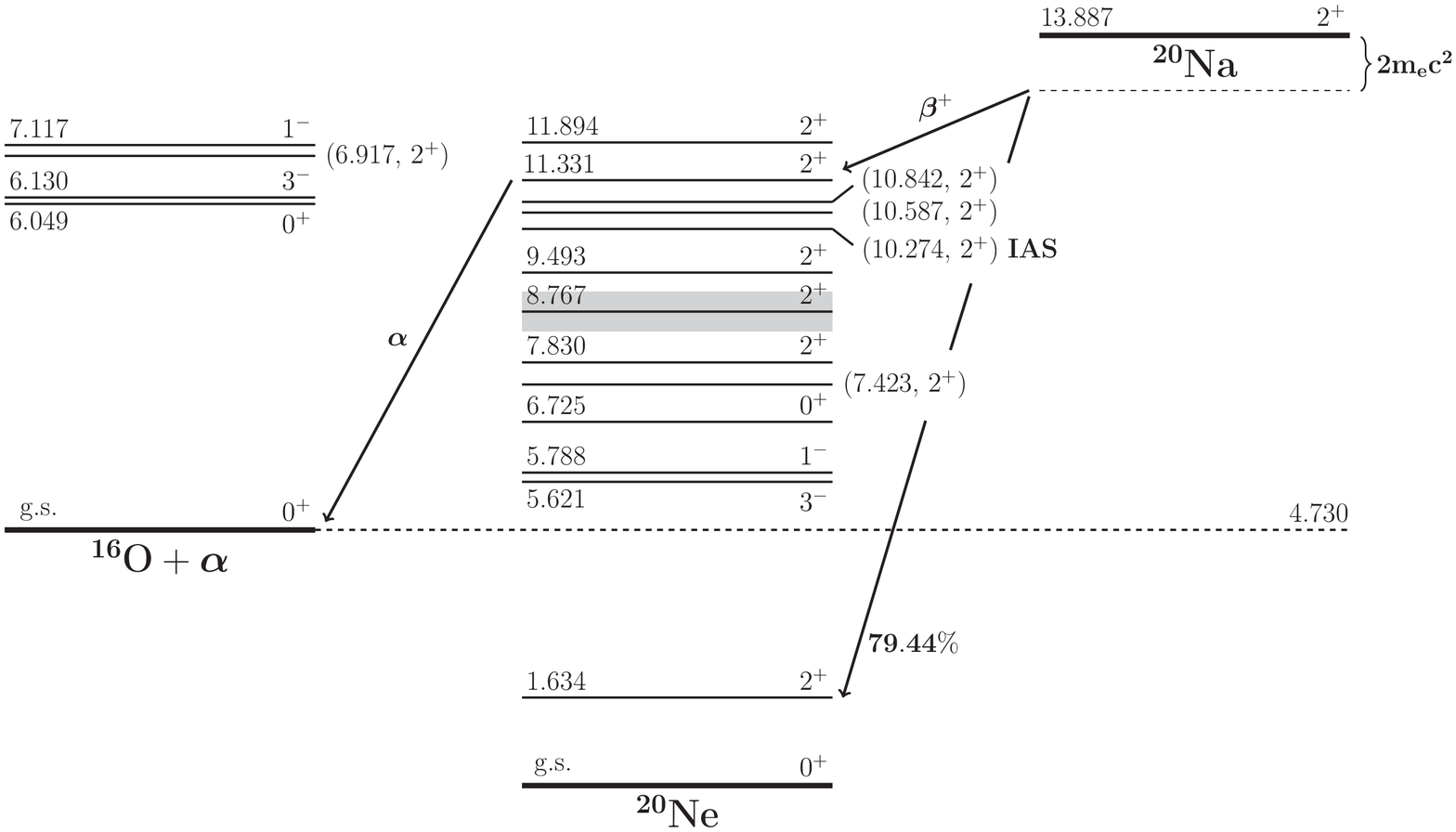}
\caption{\label{fig:scheme}Decay scheme for $^{20}$Na. Details of
  several rare decays are given in Table \protect\ref{tb : decay
    scheme}. Gamma rays in $^{20}$Ne are
  not marked.}
\end{figure*}

The serendipitous improvements to the $\beta\alpha$ part of the decay
scheme of $^{20}$Na (see Fig.~\ref{fig:scheme}) presented here fall in
two groups.

By measuring the $\alpha$ particle and the $^{16}$O recoil ion in
coincidence, we were able to identify three new $\beta$-delayed
$\alpha$ groups below 1.5~MeV that, in the singles spectrum, are
buried under the up to five orders of magnitude more intense $^{16}$O
recoil groups. The two lowest groups could be due to first forbidden
transition (they arise from known negative parity states in
$^{20}$Ne), but at least the lowest one has an important contribution
from $\beta\gamma\alpha$ decays through higher lying $^{20}$Ne
states. The origin of the third group is less certain, but it could be
due to a combined EC-alpha-gamma and beta-alpha-gamma decay through excited states in $^{16}$O. A
fourth low-energy group is also due to $\beta\gamma\alpha$ decays.

The high level of statistics gathered in the present experiment
allowed for an $R$-matrix analysis of the interference features seen
in the $\beta$-delayed $\alpha$ spectrum around 3~MeV and 5~MeV. The
spectrum turns out to be describable in terms of levels that all have
been seen in reaction experiments. Compared to earlier $\beta$-decay
studies we remove three suggested levels, but confirm conclusively
$\beta$ feeding to the broad state close to 9~MeV excitation energy in
$^{20}$Ne. The large width of this state suggests a pronounced
$\alpha+{}^{16}\textrm{O}$ cluster structure; our analysis gives
improved values for its energy and width.

Most remaining questions on the $^{20}$Na decay are minor and
involve $\gamma$-decay. A dedicated study where $\alpha$-$\gamma$
coincidences were recorded may clarify the situation concerning the
observed new low-energy $\alpha$ groups.
Also of interest could be a search for population of unbound states in $^{16}$O by detections of two-alpha events with a setup similar to the one used in the present work.

\begin{acknowledgments} 
This work has been supported by the EU 6th Framework programme "Integrating Infrastructure Initiative - Transnational Access",
Contract Number: 506065 (EURONS, JRA TRAPSPEC), by the Academy of Finland under the Finnish Centre of Excellence Programme 2006-2011 (Project No. 213503, Nuclear and Accelerator Based Physics Programme at JYFL), by the Nordic Infrastructure Project (NordForsk Project No. 070315) and by the
Spanish Funding Agency MICINN on the project FPA2009-07387
O.S.K. acknowledges the support from the Villum Kahn Rasmussen
Foundation, and A.S. acknowledges the support from the Jenny and Antti Wihuri Foundation.
\end{acknowledgments}

\bibliographystyle{natbib}

\end{document}

%% file: sec3_lowenergy.tex
\section{Data analysis and results}

\begin{figure*}
  \centering
  \includegraphics[width=0.9\linewidth,clip=true,trim=30 275 20 20]{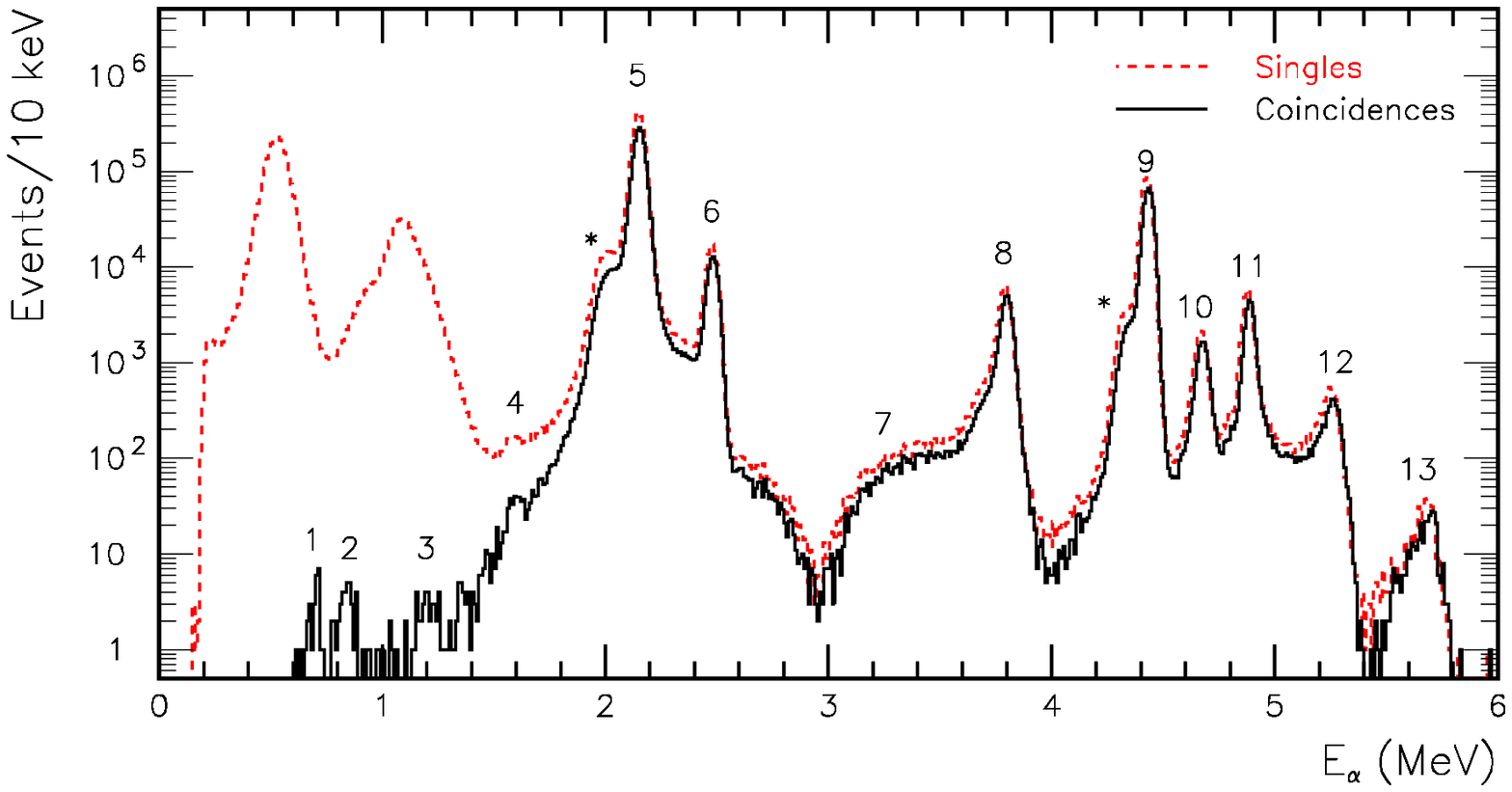}
  \caption{(Color online) $\beta$-delayed $\alpha$ spectrum of $^{20}$Na. The solid
    line (black) is the $\alpha+{}^{16}\textrm{O}$ coincidence
    spectrum. The dashed line (red) is the singles spectrum. The
    $^{16}$O recoil lines visible at 540 and 1\,110~keV, corresponds to the two most intense
    $\alpha$ lines (5 and 9). Asterisks indicate satellite peaks
    due to the additional energy lost by $\alpha$~particles striking
    the detector aluminum grid.}
  \label{fig : alpha spectra}
\end{figure*}

The singles- and coincidence $\alpha$-spectra, summed over all four
detectors, obtained in the second run are shown in Fig.~\ref{fig : alpha spectra}. 
The $^{16}$O recoil lines corresponding to the two most intense
$\alpha$ lines (5 and 9) are clearly visible in the singles spectrum
at 540 and 1\,110~keV, i.e. one-fourth of the $\alpha$-particle energies.
Only coincidence events which fulfill the requirement $E_{^{16}\textrm{O}} <
\tfrac{1}{4} E_{\alpha}+75\textrm{~keV}$ are used to generate the coincidence
spectrum. This cut greatly reduces the response tails. In particular,
the response tail of the most intense $\alpha$ line (5) is greatly
reduced, clearly exposing the new $\alpha$ lines below 1.5~MeV.
Satellite peaks, caused by additional energy loss of $\alpha$
particles striking the aluminum grid covering 4--5\% of the
detector surface, are visible on the low-energy
flank of the two most intense $\alpha$ lines and are marked
with asterisks.

\subsection{New low-energy lines}

This section is concerned with $\alpha$ lines 1--4; lines 1--3 have
not been observed previously.
As we shall argue, lines 1 and 2 may be identified with
$\alpha$ decays of the $3^-, \, T=0$ state at 5\,621.4(17)~keV and the 
$1^-, \, T=0$ state at 5\,787.7(26)~keV. The corresponding $\alpha$-particle 
energies are 713~keV and 846~keV and the $^{16}$O energies are
178~keV and 212~keV. 
The two new $\alpha$ lines are only seen in DSSSD~3, the
reason being that DSSSD~2, which is placed opposite of DSSSD~3, is the only
detector with sufficiently low ADC cutoffs to detect the coincident $^{16}$O ions.
The $^{16}$O ions lose a substantial part of their energy
in the carbon foil and in the detector dead layer. At average
implantation depth and $45^{\circ}$ exit angle, the energy loss is
approximately~50~keV. In the active volume of the detector another 30~keV is
lost to non-ionizing processes, thus reducing the detectable $^{16}$O energies to
approximately 100 and 130~keV, respectively, which is comparable to
the typical ADC cutoff in DSSSD~2. 
Below, we discuss the consequences this has for the low-energy
$\alpha$ lines.

\subsubsection{Reduced detection efficiency and energy shift}
\begin{figure}
\centering
\includegraphics[width=1.0\columnwidth,clip=true,trim=0 60 40
100]{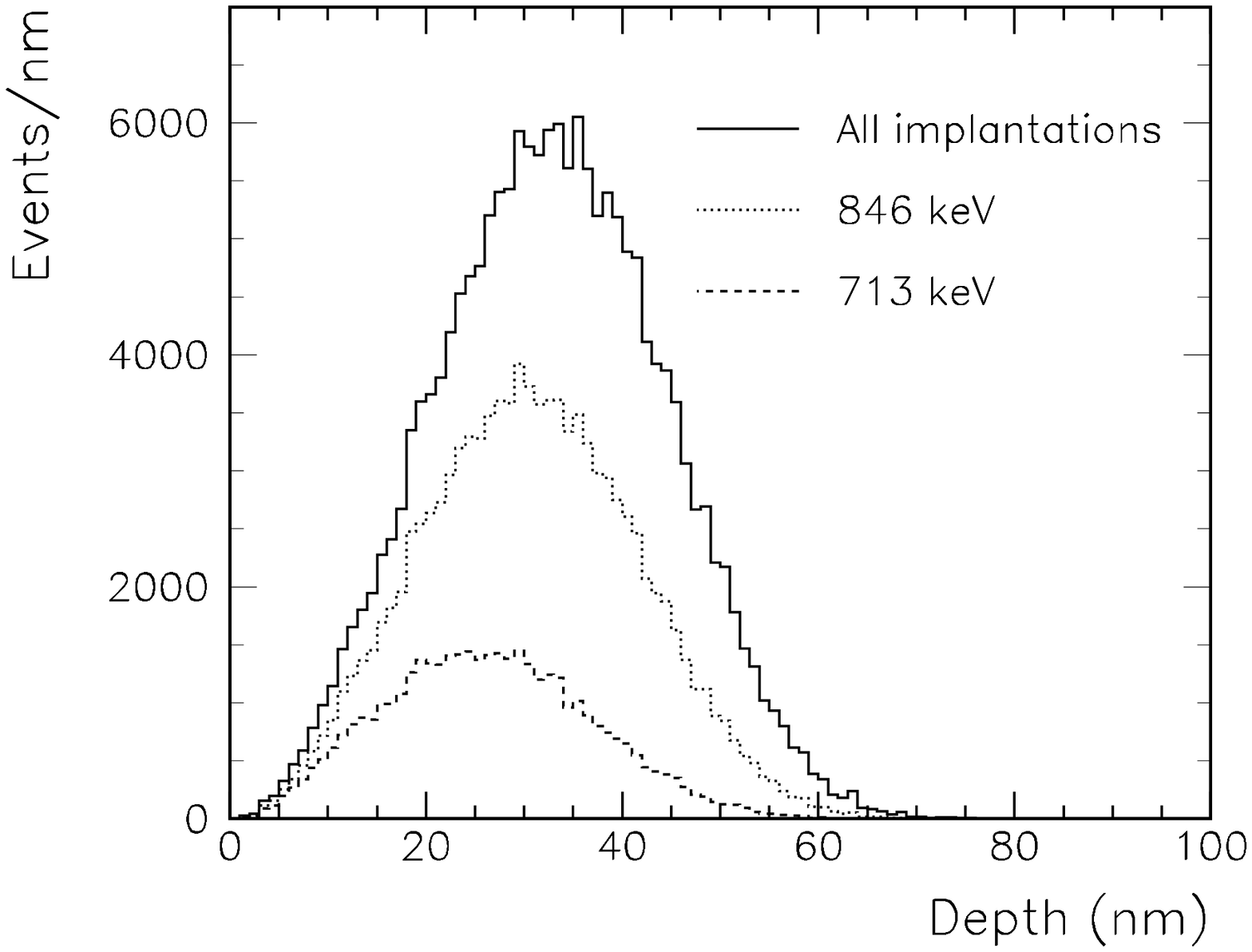}
\caption{Implantation-depth distributions obtained from Monte Carlo
  simulations showing the reduction in coincidence-detection
  efficiency and the apparent shift toward shallow implantation for the $\alpha$
  lines at 713~keV and 846~keV.}
\label{fig : implant}
\end{figure}

The implantation-depth distribution of 20~keV $^{20}$Na ions in
carbon obtained from a TRIM simulation~\cite{SRIM} is shown by the
solid line in Fig.~\ref{fig : implant}.
The large width of the distribution causes a significant broadening of
the $^{16}$O energies ($\approx 30$~keV FWHM), with shallow implantations giving the largest
$^{16}$O energies because DSSSD~2 is located upstream with respect to
the foil. The recoil shift discussed in Section~\ref{sec : recoil}
also contributes to the broadening ($\approx 30$~keV FWHM). 
Owing to the ADC cutoffs there exists a preference for
shallow implantation and positive recoil shift, implying a
larger-than-average energy loss and a
negative recoil shift for the $\alpha$ particles. 

To determine the reduction in coincidence-detection efficiency and the shift in the apparent
$\alpha$-particle energy due to the effects discussed above, we perform Monte Carlo simulations~\cite{kirsebom11_8B}.
The reduction in detection efficiency and the preference for shallow
implantation is clearly seen in Fig.~\ref{fig : implant}, which shows
implantation-depth distributions predicted by the Monte Carlo simulations: If all
implantations are considered, the solid distribution results. If, on
the other hand, we only consider implantations where both the $\alpha$
particle and the $^{16}$O ion are detected, the dotted and
the dashed distributions result.

In the first run, $\alpha$ lines 1 and 2 are observed at $699(4)$~keV and
$849(8)$~keV. The apparent shift in $\alpha$ energy determined from the simulations
are $-14(5)$~keV and $-4(4)$~keV. The systematic uncertainty on the energy shifts was
estimated by performing simulations with all ADC cutoffs
raised and lowered by 10~keV and with the detector dead layer increased
and decreased by 10\%. Correcting the observed energies for the shifts, we obtain
$713(6)$~keV and $853(9)$~keV.
In the second run, $\alpha$ lines 1 and 2 are observed at $698(4)$~keV and
$838(5)$~keV. The $\alpha$ energy shifts determined from the simulations
are $-16(5)$~keV and $-6(4)$~keV. Compared to the first run, the shifts in $\alpha$ energy are
slightly larger. This is due to the accumulation of material on the
foil between the two runs, resulting in a 20\%
increase in foil thickness.
Correcting the observed energies for the bias, we obtain
$714(6)$~keV and $844(9)$~keV.
Taking the average of the two runs, we obtain $714(4)$~keV and
$847(5)$~keV in excellent agreement with the literature values, 713.2(14)~keV
and 846(2)~keV.

\subsubsection{Relative intensities}

The determination of the intensities of the low-energy $\alpha$ lines relative to the most intense $\alpha$ line is presented in
Table~\ref{tb : lowe}.
\begin{table}
\caption{\label{tb : lowe} Relative intensities of $\alpha$ lines
  1--5 and 9. $N$ is the number of
coincidence events observed in the first (top) and second (bottom) run 
with the $\alpha$ particle detected in DSSSD~3 and the $^{16}$O ion
detected in DSSSD~2. 
$\varepsilon$ is the coincidence-detection
efficiency determined from Monte Carlo simulations (arbitrarily fixed to 100\%
for line 9). 
$I$ is the efficiency-corrected intensity relative to that of line 5 with the
statistical uncertainties given in parentheses. The average of the
two runs is given in the next-to-last column and the systematic uncertainty
on the intensity is given in the last column.} 
\renewcommand{\thefootnote}{\alph{footnote}}
\begin{tabular}{cccccc} 
\hline\hline
Line\T\B   &  $N$  &  $\varepsilon$ (\%)  &  \multicolumn{2}{c}{$I$
  (\%)} 
& Syst.~uncert. \\
\hline

\multirow{2}{*}{$1$}  & 23\T & 24 & $0.043(9)$ & \multirow{2}{*}{0.028(4)}
& \multirow{2}{*}{$_{-0.013}^{+0.04}$} \\  
                             & 25\B & 17 & $0.023(5)$ & & \\  

\multirow{2}{*}{$2$}  & 15\T & 65 & $0.010(3)$ & \multirow{2}{*}{0.0101(14)}
& \multirow{2}{*}{$_{-0.0016}^{+0.003}$} \\  
                             & 37\B & 57 & $0.0100(16)$ & & \\  

\multirow{2}{*}{$3$}  & 22\T & 89.7 & 0.010(4) & \multirow{2}{*}{0.006(2)}
& \multirow{2}{*}{$\pm 0.0002$} \\  
                             & 29\B & 87.7 &0.005(2) & & \\

\multirow{2}{*}{$4$}  & 18\T & 93.9 & $0.009(2)$ & \multirow{2}{*}{0.0083(10)}
& \multirow{2}{*}{$\pm 0.0002$} \\  
                             & 50\B & 93.8 & $0.0082(12)$ & & \\

\multirow{2}{*}{$5$}  & $2.139\times 10^5$\T & 95.6 & 100 & \multirow{2}{*}{100}
& \multirow{2}{*}{$_{\cdots}^{\cdots}$} \\  
                             & $6.182\times 10^5$\B & 95.4 & 100
                             & & \\  

\multirow{2}{*}{$9$}  & $3.986\times 10^4$\T & 100 & $17.81(10)$ & \multirow{2}{*}{17.85(5)}
& \multirow{2}{*}{$\pm 0.3$} \\  
                             & $1.157\times 10^5$ &\B 100 & $17.86(6)$
                             & & \\  

\hline
\end{tabular}
\end{table}
The number of coincidence events observed, $N$, with the $\alpha$
particle detected in DSSSD~3 and the $^{16}$O ion detected in DSSSD~2,
is given in the second column. 
The coincidence-detection efficiency, $\varepsilon$, determined from
Monte Carlo simulations and arbitrarily fixed to 100\%
for line 9, is given in the third column.
The efficiency-corrected intensity, $I$, relative to that of line 5, 
is given in the fourth column, and the average of the two
runs is given in the fifth column.
Finally, the systematic uncertainty
on the intensity
is given in the last column.
The results of the two runs are generally
in good agreement. Note the large systematic uncertainty on line 1.

For lines 1 and 2, the systematic uncertainty is dominated by the
uncertainty on the ADC cutoffs and the dead layer thickness.
For lines 3--5 and 9, the systematic uncertainty is dominated by
uncertainties in the modeling of the $^{16}$O ions as they strike the aluminum grid that covers 4--5\% of
the detector surface.

\subsubsection{Origin of the low-energy lines} 

For easier reference, our suggested decay sequences are summarized in
Table \ref{tb : decay scheme}. The detailed arguments are as follows.

\begin{table}
\caption{\label{tb : decay scheme} Overview of the rare decay
  sequences inferred from the observation of the new low-energy lines.} 
\renewcommand{\thefootnote}{\alph{footnote}}
\begin{tabular}{clcc} 
\hline\hline
Line\T\B   &  \multicolumn{1}{c}{Decay sequence}  &  $b_{\beta}$ (\%)  &  $\Gamma_{\gamma}/\Gamma$ (\%)\\
\hline

1\T & 
$^{20}\textrm{Ne}(5.621) 
\stackrel{\alpha}{\rightarrow}
{}^{16}\textrm{O}_{\text{gs}}$ & 
$<0.007$ & $\dots$\\
& 
$^{20}\textrm{Ne}(9.873)
\stackrel{\gamma}{\rightarrow} {}^{20}\textrm{Ne}(5.621) 
\stackrel{\alpha}{\rightarrow}
{}^{16}\textrm{O}_{\text{gs}}$ & 
$0.028(14)$\footnotemark[1] & $\approx 7$\footnotemark[1]\\
&
$^{20}\textrm{Ne}(10.274)
\stackrel{\gamma}{\rightarrow} {}^{20}\textrm{Ne}(5.621) 
\stackrel{\alpha}{\rightarrow}
{}^{16}\textrm{O}_{\text{gs}}$ & 
$2.877(42)$\footnotemark[1] & $0.084(17)$\footnotemark[2]\B\\

2\T & 
$^{20}\textrm{Ne}(5.788) 
\stackrel{\alpha}{\rightarrow}
{}^{16}\textrm{O}_{\text{gs}}$ & 
$0.0016(5)$ & $\dots$\B \\

3\T & 
$^{20}\textrm{Ne}(12.39) 
\stackrel{\alpha}{\rightarrow}
{}^{16}\textrm{O}(6.130)
\stackrel{\gamma}{\rightarrow}
{}^{16}\textrm{O}_{\text{gs}}$ & 
$0.0010(3)$\footnotemark[3] & $\dots$\B \\

4\T & 
$^{20}\textrm{Ne}(10.274)
\stackrel{\gamma}{\rightarrow} {}^{20}\textrm{Ne}(6.725) 
\stackrel{\alpha}{\rightarrow}
{}^{16}\textrm{O}_{\text{gs}}$ & 
$2.877(42)$\footnotemark[1] & $0.046(6)$\footnotemark[4]\\
&
$^{20}\textrm{Ne}(11.262)
\stackrel{\gamma}{\rightarrow} {}^{20}\textrm{Ne}(6.725) 
\stackrel{\alpha}{\rightarrow}
{}^{16}\textrm{O}_{\text{gs}}$ & 
$0.205(26)$\footnotemark[1] & $6.5(8)$\footnotemark[4]\B\\

\hline
\footnotetext[1]{Value from Ref.~\cite{tunl_A20}}
\footnotetext[2]{Using $\Gamma_{\gamma}=0.097(14)$~eV~\cite{fifield77}
  and $\Gamma=116(17)$~eV determined in the present work.}
\footnotetext[3]{The ratio of the phase space available for $\beta$ decay and electron-capture decay (EC) to the 12\,390~keV state suggests that this would be a 'mixed' transition with EC/$\beta = 0.16$.}
\footnotetext[4]{Branching ratio inferred under the assumption that
  the decay sequence accounts for all the observed events.}
\end{tabular}
\end{table}

\paragraph{Lines 1 and 2.} The energies of the two lowest-lying
$\alpha$ lines are consistent with $\alpha$ decay of the $3^-, \, T=0$
state at 5\,621~keV and the $1^-, \, T=0$ state at 5\,788~keV to
the ground state of $^{16}$O. 
The intensities are consistent with first-forbidden $\beta$
transitions. 
However, it is possible that allowed $\beta$ transitions to higher lying states followed
by $\gamma$ decay to the 5\,621~keV and 5\,788~keV states also contribute. In
fact, the IAS is known to decay to the 5\,621~keV state with a partial
$\gamma$ width of $\Gamma_{\gamma_1} =
0.097(14)$~eV~\cite{fifield77}. 
The $\alpha$ width of the IAS has been determined to be
$\Gamma_{\alpha}=116(20)$~eV~\cite{ingalls76}.
It follows that 
$\Gamma_{\gamma_1}/\Gamma_{\alpha}=8.4(19)\times 10^{-4}$.
In comparison, the efficiency-corrected relative intensity determined
from the present data is $I_1/I_9=1.6(2)^{+2.2}_{-0.7}\times 10^{-3}$,
 implying that decays through the IAS account for approximately 20--100\% of the observed intensity of $\alpha$ line 1; the wide range reflects the large uncertainty on the efficiency correction.
Furthermore the $3^+$ state at 9\,870~keV is known \cite{clifford89} to
be fed in $\beta$-decay with a branching ratio 0.028(14) and to
$\gamma$-decay to the 5\,621~keV state with a branching ratio about
0.07 \cite{tunl_A20}, thereby giving a contribution of roughly the same magnitude
as decays through the IAS.
Note that the efficiency correction assumes a first-forbidden $\beta$
transition to the 5\,621~keV state. However, since the combined recoil
caused by a $\beta$ transition to the IAS followed by a $\gamma$
transition to the $5\,621$~keV state is similar in magnitude, the two
decay modes should have similar efficiency corrections.
We note that $\gamma$ transitions to the 5\,788~keV state have not been
observed~\cite{fifield77}.

\paragraph{Line 3.} The third $\alpha$ line is located at $E_{\alpha}=1\,220(30)$~keV, which
corresponds to an excitation energy of 6\,260(40)~keV in $^{20}$Ne,
assuming that the $\alpha$ decay proceeds to the ground state of $^{16}$O. No
state is known in $^{20}$Ne at this energy. In fact, the observed
width of line 3 is incompatible with such a hypothesis:
Assuming a reduced $\alpha$ width equal to the Wigner
limit~\cite{teichmann52}, one obtains an $s$-wave $\alpha$ width of only
7.9~keV due to the small relative
energy.\footnote{This calculation assumes a `standard' channel radius of
  $a=1.4\textrm{ fm}\times (4^{1/3}+16^{1/3}) = 5.8$~fm.} Since no
other particle-decay channels are open, this represents an upper limit
on the total width of the state.
However, the observed width of line 3 is $150(30)$~keV, well above the estimated
experimental resolution of $\approx 35$~keV. Converting the energy
scale from $\alpha$-energy to excitation energy, this corresponds to a width
of $\Gamma = 190(40)$~keV, well above the Wigner limit. 

\begin{figure}
\centering
\includegraphics[width=1.0\columnwidth,clip=true,trim=0 80 50
100]{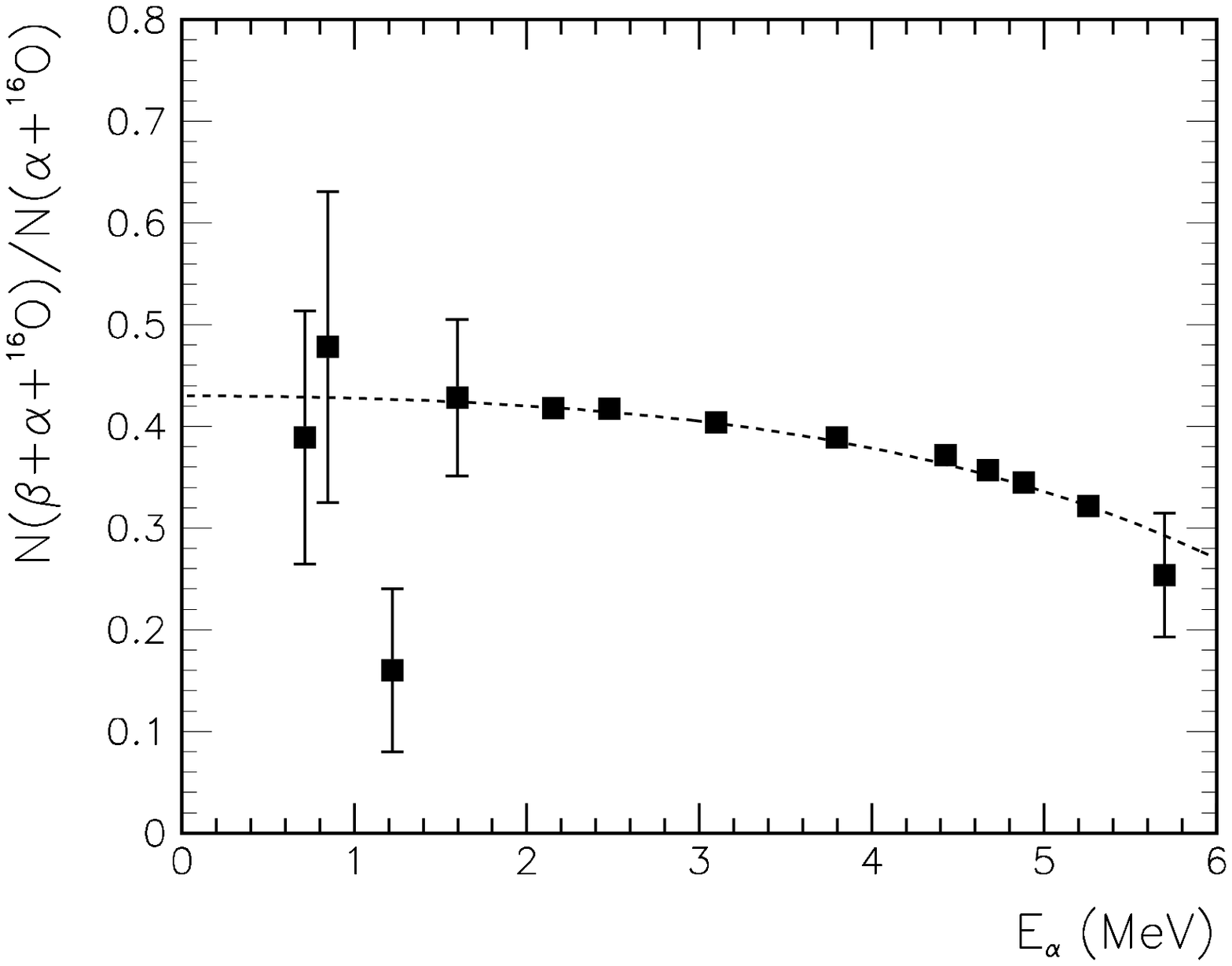}
\caption{Ratio of triple coincidences ($\beta+\alpha+{}^{16}$O) to double
  coincidences ($\alpha+^{16}$O). The dashed curve is only meant to
  guide the eye. The error bars show the statistical uncertainty.}
\label{fig : beta}
\end{figure}

$\beta$ particles were detected in the 1.5~mm thick silicon detectors
placed behind the DSSSDs. The ratio of triple coincidences
($\beta+\alpha+{}^{16}$O) to double coincidences ($\alpha+^{16}$O) is
shown in Fig.~\ref{fig : beta}. This `$\beta$ ratio' displays a clear
systematic dependence on $\alpha$ energy except for line 3, which has
a very small $\beta$ ratio, consistent with zero at the 2$\sigma$
level. This behaviour is suggestive of electron capture (EC). We have
identified one possible decay sequence: The $1^+$, 13\,308~keV state in
$^{20}$Ne is only fed in EC decay. It $\alpha$ decays to the
$3^-$/$2^+$/$1^-$ states in $^{16}$O (cf.~Fig.~\ref{fig:scheme}) with
partial widths of 700/19/190 in units of eV, emitting $\alpha$
particles with energies of 1\,960/1\,330/1\,170 in units of keV. The main
$\alpha$ line at 1\,960~keV would be buried under the tails of
higher-lying $\alpha$ lines. However, the observed intensity of line 3
implies a $\log ft$ value of 2.3, an extremely unlikely possibility.
We have searched for other potential candidates among the known states
in $^{20}$Ne, but have failed to find any. The origin of line 3 thus
remains mysterious. 

We note, however, that the sd-shell model
calculations of Ref.~\cite{Brown85} predict both a $3^+$ state at 11\,400~keV and a $1^+$ state at 12\,200~keV to be populated in $\beta$ decay.
None of these states have been observed in $\beta$ decay yet, but $\beta$
decay to the $1^+$ state with the calculated strength followed by
$\alpha$ decay to the $3^-$ state in $^{16}$O would explain the
properties of line 3, provided the $1^+$ state lies at the slightly
higher energy of 12\,390~keV.
At this elevated energy $\beta$ decay will still be favored over EC decay with a phase-space ratio of EC/$\beta = 0.16$. It would thus appear that the proposed decay sequence is inconsistent with the anomalous low $\beta$ ratio of line 3. However, the parameter driving the systematic behavior seen in Fig. 4 is the $\beta$ energy, not the $\alpha$ energy. In the proposed decay sequence the $\beta$ energy is much smaller than the $\alpha$ energy would indicate because the decay feeds an \textit{excited} state in $^{16}$O; if the decay were to feed the ground state the $\alpha$ energy would be 6\,130~keV. Indeed, the $\beta$ ratio of line~3 is in good agreement with the observed systematics at 6\,130~keV.

We note that the $\alpha$ threshold in $^{16}$O is only 1\,MeV above the $3^-$ state and that, in principle, population of unbound states in $^{16}$O is possible in the decay of $^{20}$Na. We have searched for such $\beta$-delayed two-$\alpha$ events, but found no significant signal above background.

\paragraph{Line 4.}
In Ref.~\cite{clifford89} it was suggested that line~4 should be
attributed to the $\alpha$ decay of the $0^+$, $T=0$ state in $^{20}$Ne at
6\,725(5)~keV. The excitation energy deduced from the present data
is 6\,716(6)~keV, thus supporting this interpretation.
The intensity is a factor of two below the intensity obtained in
Ref.~\cite{clifford89}. We are unable to account for this discrepancy.
As argued in Ref.~\cite{clifford89}, the direct population of a $0^+$
state requires a second-forbidden transition, which results in a
much smaller intensity than the intensity actually measured. 
Two alternative explanations were offered: 
(i) a $\beta$ branch feeding a hypothetical state at 12\,758~keV,
which $\alpha$ decays to the $0^+$ first-excited state in $^{16}$O at
6\,049~keV, and
(ii) a weak $\beta$ branch feeding a hypothetical state around
$11\,000$~keV in $^{20}$Ne, which decays mainly by $\gamma$-ray emission
to the 6\,725~keV state. 
It was already noted in Ref.~\cite{clifford89} that alternative (i) is unlikely to be
the correct explanation because the $\beta$ branch would have a very
small $\log ft$ value of 3.28. 
We can fully rule out alternative (i): Since the 6\,049~keV state in
$^{16}$O decays by $e^+e^-$ emission, the
$\beta$ ratio should be significantly higher (roughly a
factor of 3) for $\alpha$ line 4 compared to all other $\alpha$ lines. 
Fig.~\ref{fig : beta} clearly shows that this is not the case.
We cannot confirm nor rule out alternative (ii). 
Instead, we offer a third and fourth alternative: (iii) a weak $\gamma$ branch
from the IAS to the 6\,725~keV state and (iv) a weak $\gamma$ branch
from the $1^+$, $T=1$ state at 11\,262.3(19)~keV to the 6\,725~keV
state. 

We first consider alternative (iii): Based on the observed
intensities, we can easily estimate the width of the proposed $\gamma$
branch,
\begin{equation}
\Gamma_{\gamma_4} = \Gamma_{\alpha} \frac{I_4}{I_9} =
0.054(11)\textrm{ eV}\; ,
\end{equation}
where $I_4$ and $I_9$ are the efficiency-corrected relative
intensities given in Table~\ref{tb : lowe} and $\Gamma_{\alpha} =
116(20)$~eV is the $\alpha$ width of the IAS~\cite{ingalls76}.
The least intense $\gamma$ branch observed by Fifield {\it et al.}\ was
that to the 4\,967~keV with a
partial width of 0.060(8)~eV~\cite{fifield77}. It is difficult to
judge from their Fig.~2 whether a 0.054(11)~eV branch to the 6\,725~keV
state could have gone unnoticed, but it appears possible, especially
taking into account that the 6\,725~keV state is fairly wide
($\Gamma=19.0(9)$~keV).

Next, we consider alternative (iv):
The 11\,262~keV state is fed with a $\beta$-branching ratio of $b_{\beta}(11.262)=0.205(26)\%$
corresponding to a $\log ft$ value of 3.72~\cite{clifford89}. 
$\alpha$ decay from the 11\,262~keV state is parity
forbidden\footnote{The parity-forbidden $\alpha$-decay width has been
  determined to be $42(20)\times 10^{-6}$~eV~\cite{fifield83}.} so it almost
exclusively decays by $\gamma$ emission. It is known to decay to the
ground state and the first-excited state with relative intensities of
84(5)\% and 16(5)\%, respectively~\cite{berg83}, and the partial width to the
ground state has been determined to be 11(2)~eV~\cite{bendel71}. 
Based on the observed intensity of line 4, we can estimate the width
of the proposed $\gamma$ branch to be 
\begin{equation}\label{eq:1+}
\Gamma_{\gamma_4} = I_4\Gamma_{\gamma}
\frac{b_{\beta}(7.421)}{b_{\beta}(11.262)} = 0.086\textrm{ eV}\; ,
\end{equation}
where $I_4$ is the efficiency-corrected relative
intensity given in Table~\ref{tb : lowe}, $\Gamma_{\gamma}=13.3(2)$~eV is
the total width of the 11\,262~keV state and $b_{\beta}(7.421)= 15.96(22)\%$
is the $\beta$-branching ratio to the 7\,421~keV state.
The deduced width corresponds to an M1 transition
strength of 0.043~W.u., which is a reasonable value. Furthermore, the
branch is far smaller than the two known branches, so it could easily
have been missed in the study of Berg {\it et al.}~\cite{berg83};
unfortunately, their $\gamma$ spectrum only extends down to 6~MeV.

\subsection{Energies and relative intensities}

The literature data on the energies and relative intensities of the
$\beta$-delayed $\alpha$ lines of $^{20}$Na are from
Refs.~\cite{torgerson73, honkanen81, clifford89, huang97,
  hyldegaard09}. In Table~\ref{tb : results}, we compare the present
data to that of Refs.~\cite{clifford89, huang97, hyldegaard09}.  The
lines are numbered as in our spectrum and letters are used to denote lines
quoted in other papers, but not seen by us (as argued below most of these are due to
misinterpretations).
\begin{table*}
\caption{Comparison of present energies and
  relative intensities to previous data.} \label{tb : results}
\renewcommand{\thefootnote}{\alph{footnote}}
\begin{tabular}{cccccccc} 
\hline\hline
\multirow{2}{*}{Line}\T   &  \multicolumn{3}{c}{$E$~(keV)}  &  
\multicolumn{4}{c}{$I_4$~(\%)}  \\
 &  Present\B  &  Ref.~\cite{clifford89}  &  Ref.~\cite{huang97}  &
 Present\B  &  Ref.~\cite{clifford89}  &  Ref.~\cite{huang97} &
 Ref. \cite{hyldegaard09} \\
\hline

1  & $714(4)$ & $\cdots$ & \multirow{2}{*}{$\sim 780$\footnotemark[1]}
& $0.028(4)(_{-13}^{+40})$ & $\cdots$ &
\multirow{2}{*}{$\sim 1.4$\footnotemark[1]} &
\multirow{2}{*}{ \ } \\  

2  & $847(5)$ & $\cdots$ & & $0.0101(14)(_{-16}^{+30})$ & $\cdots$ & & \\  

3  & $1220(30)$ & $\cdots$ & $\cdots$ & $0.006(2)$ & $\cdots$ & $\cdots$ &  \\  
4  & $1\, 589(5)$ & $1\, 580(40)$ & $\cdots$ & $0.0083(10)(2)$ & $0.020(4)$ &
 $\cdots$ & \\
5  & $2\, 153.2(10)$\footnotemark[2] & $2\, 150.4$\footnotemark[2] &
 $2\, 150.4$\footnotemark[2] & $100$ & $100$ & $100$ & $100$ \\  
6  &  & $2\, 479.6(21)$ & $2\, 483.5(25)$ & & $3.7(4)$ & $3.91(24)$ & $3.7(3) $ \\  
A   &  & $2\, 659(7)$ & $2\, 756(5)$ & & $0.074(6)$ & $0.072$ & \\  
7  &  & $3\, 570(25)$ & $3\, 325(13)$ & & $0.39(4)$ & $0.052$ & \\  
8  &  & $3\, 799(3)$ & $3\, 803.0(25)$ & & $1.510(27)$ & $1.55(5)$ & $1.6(2)$ \\  
9  & $4\,433.8(15)$\footnotemark[2] & $4\, 432.2$\footnotemark[2] &
 $4\, 432.2$\footnotemark[2] & $17.85(5)(30)$ & $17.31(9)$ & $17.83(27)$ & $16.3(3)$ \\ 
10  &  & $4\, 675(3)$ & $4\, 674.6(21)$ & & $0.553(15)$ & $0.478(24)$ & $0.45(6)$ \\  
11  &  & $4\, 885(3)$ & $4\, 884.4(25)$ & & $1.09(3)$ & $1.15(4)$ & $0.91(8)$ \\  
B   &  & $4\, 966(7)$ & $4\, 930(6)$ & & $0.075(9)$ & $0.128$ & \\  
C   &  & $5\, 106(7)$ & $5\, 222(47)$ & & $0.055(7)$ & $0.044$ & \\  
12  &  & $5\, 249(4)$ & $5\, 253.5(23)$ & & $0.165(11)$ & $0.130(9)$ & \\  
13  &  & $5\, 698(6)$ & $5\, 691(4)$ & & $0.010(2)$ & $0.012(2)$ & \\  
D   & $\cdots$ & $\cdots$ & $5\, 896(6)$ & $\cdots$ & $\cdots$ & $0.002$ & \\  
\hline
\end{tabular}
\footnotetext[1]{The evidence presented in Ref.~\cite{huang97} for the observation of this $\alpha$ group is meager.} 
\footnotetext[2]{Used for energy calibration.} 
\end{table*}
Line D, seen by Ref.~\cite{huang97} but not by
Ref.~\cite{clifford89}, was not seen in the present study despite the
higher statistics.
The present data on the energies of lines 4--13 is consistent with the
previous data. 
The energies of the two most intense $\alpha$ lines (5 and 9) used for
the energy calibration, differ from the previous energies by a few keV. 
This is due to the updated value for the $\alpha+{}^{16}$O threshold
in $^{20}$Ne~\cite{audi03}.

The intensity of line~9 agrees with the intensity reported
in Ref.~\cite{huang97} but is slightly larger than the intensity
reported in Ref.~\cite{clifford89}. 
An older study~\cite{torgerson73} gives 17.39(11)\% in agreeement with
Ref.~\cite{clifford89}.

Only the low-energy lines and the two most intense lines are included
for the present data. The less intense lines at higher energy can be
influenced in position as well as intensity by interference effects
so we shall give our results below, when we discuss the $R$-matrix fit (Table \ref{tab:results}).

\subsection{$R$-Matrix Analysis}

We turn now to a detailed analysis of the $\beta$-delayed
$\alpha$-spectrum above 2~MeV. This part of the spectrum contains
several resonances as well as indications of interference effects at
lower intensity levels. The $R$-matrix formalism provides a suitable
framework to incorporate such effects in a fitting routine, but
reliable results can only be extracted once the experimental response
function is fully understood.


\subsubsection{R-matrix parametrization}


The $R$-matrix parametrization allows extraction of energy-, width- and
$\beta$-strength parameters for the resonance levels in question. The
parametrization offers at the same time a more proper interpretation
of the excitation energy distribution compared to earlier studies. We
use the same procedure as in Ref.~\cite{kirsebom11_8B}. The probability for $\beta$ decay to the excitation energy $E$, followed by breakup into the channel $c$, is given by
\begin{align}
w_c\left(E \right) = &\,C^2 f_\beta\left(E\right) P_c\left(E - E_\text{thres} \right) \nonumber \\ 
     &\times \sum_{x=F, GT} \biggm\vert \sum_{\lambda\mu} \widetilde{g}_{\lambda x} 
\widetilde{\gamma}_{\mu c} \widetilde{A}_{\lambda \mu} \biggm\vert^2,
\label{eq:decay_prob}
\end{align}
where the alternative parametrization of the level matrix $A_{\lambda
  \mu}$ is used, as given in Eq. 33 of Ref.~\cite{Brune2002}. In Eq. (\ref{eq:decay_prob}) the
boundary parameter $B_c$ is automatically set equal to the shift
function $S_c\left(E_\lambda \right)$ for all levels. This ensures
that the fitted parameters for the energy, $E_\lambda$, the
$\beta$ strength, $\widetilde{g}_{\lambda x}$, and reduced
$\alpha$ width, $\widetilde{\gamma}_{\lambda \mu}$, becomes the
so-called observed ones \cite{Brune2002}. The notation $x = $F, GT
refers to Fermi and Gamow-Teller decays respectively, $f_\beta$ is the
integrated phasespace available to the leptons, $C$ is a normalization
constant and $P_c$ is the penetration function with $E_\text{thres}$ being
the $\alpha$-threshold energy. The number of counts as a function of $E$ is then
\begin{align}
N\left(E\right) = \frac{NT_{1/2}}{\ln 2}\, w_c\left(E \right),
\label{eq:N(E)}
\end{align}
where $N$ is the total number of counts and $T_{1/2}$ is the half life
of $^{20}$Na \cite{tunl_A20}.

The observed $\alpha$ widths are extracted as
\begin{align}
\Gamma_\lambda^\text{o} = \frac{\sum_c 2 P_c
 \widetilde{\gamma}_{\lambda c}^2}{1+\sum_c 
 \widetilde{\gamma}_{\lambda c}^2 \tfrac{\text{d}S}{\text{d}E}\vert_{E=E_\lambda}},
\label{eq:Gamma}
\end{align} 
and the procedure for determining the $\beta$-decay matrix elements is
similar to the one in Ref. \cite{Barker1988}
\begin{align}
B_{\lambda F} = &\, C^2 g_{\lambda F}^2 \frac{B}{\ln2}\nonumber \\
 &\times \sum_c \int_0^Q P_c\left( E-E_\text{thres} \right) \biggm\vert \sum_\mu \widetilde{\gamma}_{\mu c}\widetilde{A}_{\lambda \mu} \biggm\vert^2 \text{d}E.
\label{eq:BGT}
\end{align}
Here $B = 6144.2(1.6)\,\text{s}$ \cite{Hardy2010} and $Q$ is the
$\beta$-decay $Q$-value. For $B_{\lambda GT}$ a factor of
$\tfrac{g_V^2}{g_A^2}$ enters in the denominator.  When calculating
absolute $B_{F/GT}$ values the known decay branch ($\text{BR}_\alpha =
20.05\pm0.36$ \cite{clifford89}) to $\alpha$ unbound levels in
$^{20}$Ne is used. Integration of eq. (\ref{eq:decay_prob}) (without
$C^2$) over the full energy spectrum then yields the normalisation
constant
\begin{align}
C = \left[ \frac{\text{BR}_\alpha \ln2}{T_{1/2} \int_0^Q w\left(E\right)dE} \right]^{1/2}.
\end{align}
In our analysis only the exit channel populating $^{16}$O in the
ground state has been included. Levels 11, 12 and 13 are above the
threshold for decay to exited states in $^{16}$O. Their contribution
to the reduced $\alpha$-widths have been investigated by performing a
two-channel fit including the first exited $0^+$ state in $^{16}$O at
$6\,049.4(1)\,\text{keV}$. From this test their contribution was found
to be no higher than $10^{-3}\, \sqrt{\text{keV}}$ which justifies the
single-channel analysis.

\subsubsection{Recoil Broadening}
\label{sec : recoil}

\begin{figure}
\centering
\includegraphics[width=1.0\columnwidth,clip=true,trim=30 75 80
120]{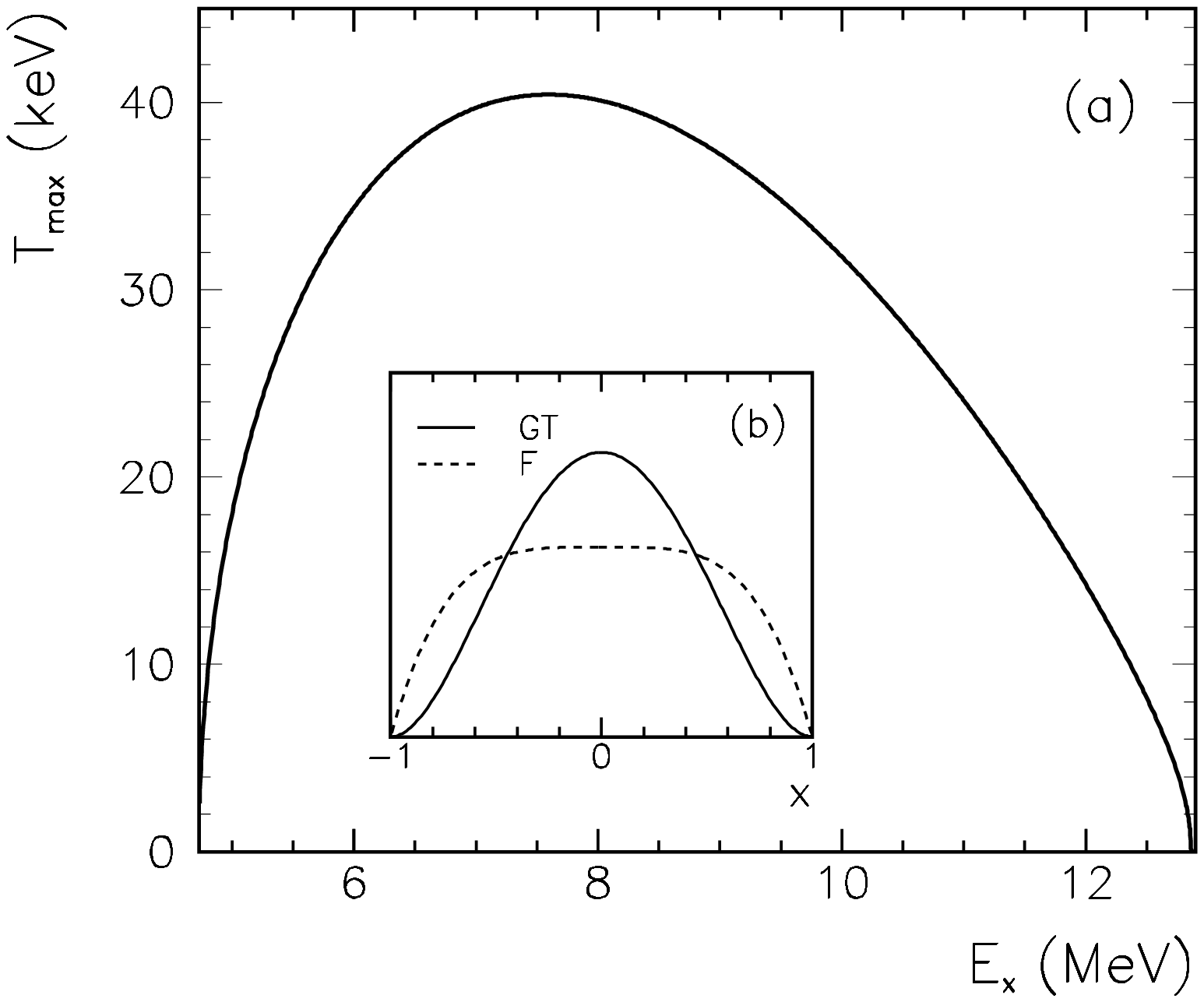}
\caption{\label{fig : recoilbroad} Recoil broadening. (a) Maximum
  recoil shift as a function of the excitation energy in
  $^{20}$Ne. (b) Fermi (F) and Gamow-Teller (GT) recoil broadening
  distributions.}
\end{figure}

If the $^{20}\textrm{Ne}^{\ast}\rightarrow \alpha+{}^{16}\textrm{O}$
breakup occurs at rest, the energies of the $\alpha$ particle and the
$^{16}$O ion are fixed by energy and
momentum conservation. The available energy is shared as $E_{^{16}\textrm{O}}/E_{\alpha} =
m_{\alpha}/m_{^{16}\textrm{O}}=0.25$.
If the breakup is preceeded by a $\beta$ decay, however, the energies
of the $\alpha$ particle and the
$^{16}$O ion are smeared out due to the recoil motion of the
$^{20}\textrm{Ne}^{\ast}$ nucleus. The recoil motion also
causes a small systematic shift of the mean energies which, however, may
be safely neglected here.
For a pure Fermi transition, the resulting energy distribution can be
approximated by \cite{bhattacharya02},
\begin{equation}\label{eq : lepton broadening (fermi)}
\rho (x) = \left\{ \begin{array}{ll} 
\frac{5}{8 T_{\textrm{max}}} (1-x^4) &, \; -1 \le x \le 1 \\
0 &, \; \vert x \vert > 1 \end{array} \right.
\end{equation}
while, for a pure Gamow-Teller transition and the spin
sequence $2^+ \rightarrow 2^+ \rightarrow 0^+$,
\begin{equation}\label{eq : lepton broadening (GT)}
\rho (x) = \left\{ \begin{array}{ll} 
\frac{15}{16 T_{\textrm{max}}} (1-2x^2+x^4) & -1 \le x \le 1 \\
0 & \vert x \vert > 1 \end{array} \right.
\end{equation}
with $x = \delta E/T_{\textrm{max}}$ where $\delta E = E_{\alpha} -
\langle E_{\alpha} \rangle$ is the shift relative to the mean
$\alpha$-particle energy and $T_{\textrm{max}}$ is the maximum shift,
given by
\begin{equation}\label{eq : Tmax Na20}
T_{\textrm{max}} = \frac{m_e}{M} \left[ 2Qm_{\alpha} c^2 (W_0^2-1)
  \frac{M c^2-m_{\alpha}c^2-Q}{M c^2-Q} \right]^{1/2} \; ,
\end{equation}
where $m_e$, $m_{\alpha}$, and $M$ are the electron, $\alpha$-particle,
and $^{20}$Ne masses; $Q = E_x - 4\, 729.84\textrm{~keV}$ and
$W_0=(E_0-E_x)/m_e c^2$ where $E_x$ is the excitation energy in $^{20}$Ne
and $E_0=13\, 376\textrm{~keV}$ is the maximum total $\beta$ energy for
decays to the ground state of $^{20}$Na.
In Fig.~\ref{fig : recoilbroad}~(a) the maximum recoil shift, $T_{\textrm{max}}$, is
shown as a function of the excitation energy in $^{20}$Ne, $E_x$. The
inset (b) shows the different shapes of the recoil broadening
distribution for Fermi (F) and Gamow-Teller (GT) transitions.

\subsubsection{Response Function}

The {\it response function} describes the distribution of energies
measured from a perfectly monochromatic source due to experimental
effects. We describe now how the response function is extracted from
the experimental data.

The second most intense line (9) in the $\beta$-delayed $\alpha$
spectrum of $^{20}$Na is found at $4\, 434$~keV and results from
transitions to the IAS in $^{20}$Ne at $10\,
273.2$~keV, the width of which is less than 
$0.3$~keV~\cite{tunl_A20}.  The transition to the IAS is
known~\cite{clifford89} to be a mixed Fermi and Gamow-Teller type
transition. However, the Gamow-Teller component is rather small so the
broadening effect is well described by Eq.~(\ref{eq : lepton
  broadening (fermi)}).  Owing to the special structure and the $T=1$
nature of the IAS, only little interference is expected with
neighboring $2^+$ states which could potentially distort the shape of
the $\alpha$ line.  The physical shape of the
IAS $\alpha$ line is therefore well understood; this 
allows for a detailed study of the modification of the shape due to
experimental effects.

We adopt a parameterization similar to the one used in Ref.~\cite{bhattacharya06} to
describe the experimental response function. It consists of a Gaussian and a Gaussian
folded through a low-energy exponential tail:
\begin{align}\label{eq : response paramatrization}
&\psi(E_0,E) \; = \; {} \, \frac{A_1}{\sqrt{2\pi \sigma}}\exp \left( \frac{\left(E-E_0\right)^2}{2\sigma^2} \right)
\nonumber \\
&{}+ \frac{A_2}{2\lambda} \exp \left( \frac{E-E_0}{\lambda} + \frac{\sigma^2}{2\lambda^2} \right) \times \textrm{erfc} \left( \frac{E-E_0+\sigma^2/ \lambda}{\sqrt{2}\sigma}
\right) \; .
\end{align}
Here $E_0$ and $E$ are the nominal and observed energies, $\lambda$
the exponential decay ``length'' and erfc the complement of the
incomplete error function. The normalization constants are $A_1 =
1/(1+r)$ and $A_2=r/(1+r)$ where $r$ gives the relative size of the second term
compared with the first one. We find $\lambda = 25.5$~keV and $r= 0.085$. In addition we allow for contributions from low energy tails of higher lying Gamow-Teller fed resonances.
 
Ions striking the aluminum grid covering $g=4.9$\% of the detector
surface experience additional energy loss compared to other ions, thus
giving rise to a satellite peak at slightly lower energy.
(The fit parameters are fairly strongly correlated so that almost
equivalent fits can be obtained with different sets of parameters; we
quote here a consistent set corresponding to the fit optimum.)
The satellite is wider than the main peak due to the
larger variation in effective thickness with angle. The final
complete response function therefore reads:
\begin{align}\label{eq : response}
\Psi(E_0,E) \; = \; &{} (1-g) \, \psi(\sigma; E_0, E) \; \nonumber \\
&{} + \; g \, \psi(\sigma_g;E_0-E_g ,E) \; ,
\end{align}
where $E_g$ is the mean energy loss in the grid, $\sigma=10.2$~keV the
Gaussian width of the main peak and $\sigma_s=25.2$~keV the
Gaussian width of the satellite peak.

\begin{figure}
\centering
\includegraphics[width=1.0\columnwidth]{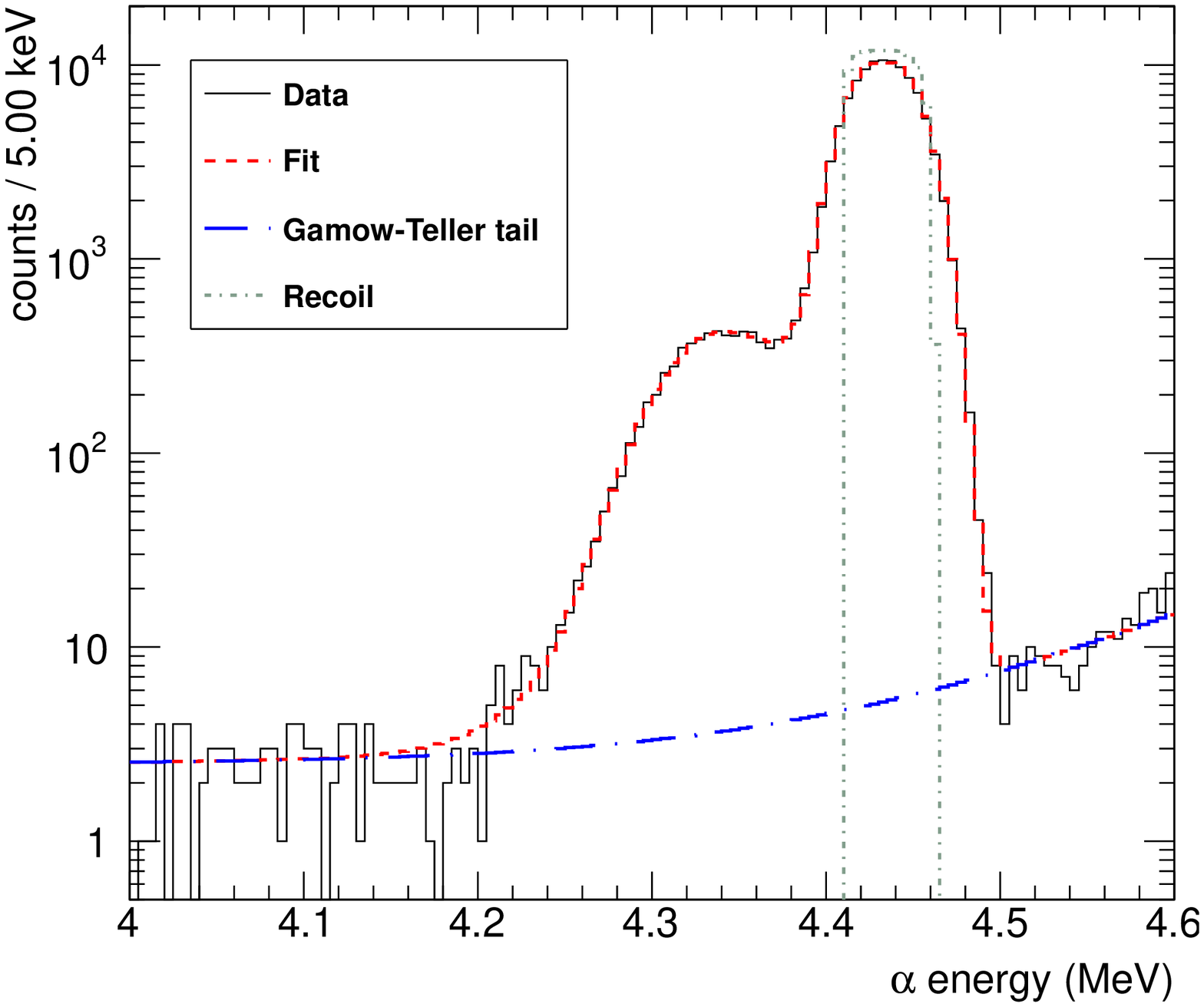}
\caption{\label{fig : response} (Color online) Singles spectrum between 4 and 4.6~MeV
  measured in DSSSD~1. The short-dash-dotted (green)
  curve shows the recoil broadening distribution, Eq.~(\ref{eq : lepton
  broadening (fermi)}). The long-dash-dotted (blue) line shows the
  contribution from GT transitions to the low-energy tails of higher-lying states, modeled as an exponential function. The dashed (red)
  curve shows the best fit to the data using Eq.~(\ref{eq : convolution}).}
\end{figure}

The shape of the IAS $\alpha$ line, including experimental
effects, is found by convolution of the response function,
Eq.~(\ref{eq : response}),
with the recoil broadening distribution, Eq.~(\ref{eq : lepton
  broadening (fermi)}), i.e.
\begin{align}\label{eq : convolution}
\frac{\textrm{d}N}{\textrm{d}E} \; = \; &{} \int \, \Psi (E_{\alpha}, E) \; \rho (x(E_{\alpha}))
\; \textrm{d}E_{\alpha} \; .
\end{align}
Fig.~\ref{fig : response} shows the line shape measured in
DSSSD~1 with the recoil broadening distribution, the response
function, and the best fit using
Eq.~(\ref{eq : convolution}) superimposed. 

The response function, $\Psi(E_0,E)$, is hereby obtained at $4\,
434$~keV. In order to determine the response function for a general
$\alpha$-particle energy, we (1) assume that the exponential tails are
independent of energy; (2) convert the energy loss in the aluminum
grid, $E_g$, into an equivalent thickness (0.60~$\mu$m) which we use
to calculate the energy loss in the grid for other $\alpha$-particle
energies; and finally (3) assume the Gaussian width of the satellite
peak, $\sigma_s$, to be proportional to $E_g$.

\subsubsection{Synthesis and Fitting}

The $ft$ values obtained in Ref.~\cite{clifford89} suggest that the
transitions to the observed resonances are first forbidden or allowed, and
hence the spin-parity could be $1^-$, $2^+$ or $3^-$. (The observation
of $\alpha$ decay from these states excludes unnatural spin-parity.)
We shall attempt first the simplest possible fit based exclusively on
allowed transitions to resonances seen earlier in reaction
experiments. All resonances will then have spin-parity $2^+$ and
interference occurs naturally.


\begin{figure*}
\includegraphics[width=1.0\textwidth]{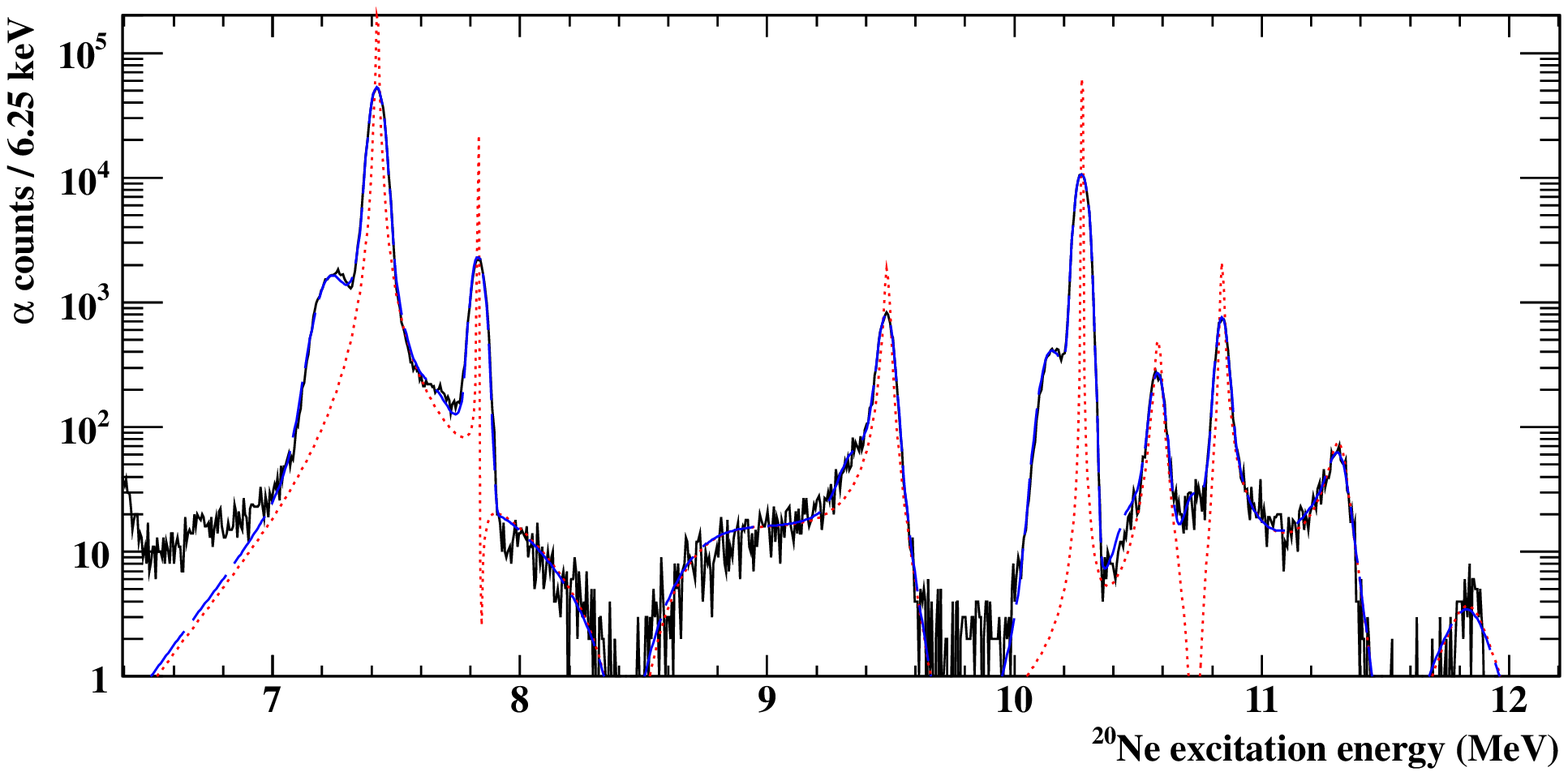}
\caption{\label{fig:fit}Best fit to the experimental data. The graph
  shows the fitted R-matrix expression (eq. \ref{eq:N(E)}) (dashed),
  the convoluted expression (long-dashed) and the data (solid).}
\end{figure*}

The delayed particle spectrum recorded in DSSSD1 during the second
$^{20}$Na run has been fitted using 
the maximum
likelihood estimator \cite{Baker1984} for the excitation-energy region 7.00--12.00~MeV. Figure
\ref{fig:fit} displays both the fitted $R$-matrix function
(eq. \ref{eq:N(E)}) and the final fit curve obtained after convolution
with the response function and the recoil broadening distribution. The
deviation between the data and the fit is largest at the satellite
shoulder of the most intense peak and dominates the rather high
$\chi^2/\text{d.o.f.} = 2.21$. This deviation is caused by
systematic errors in the propagation of the response function to other energies. The inability of the fit to describe the few events around 9.8~MeV is due to inaccurate modeling of the low-energy response tail of the IAS. The fit results are listed and compared to
literature values in Table \ref{tab:results}.

As starting point in the fitting procedure the fairly well known level
energies were fixed whereas the more uncertain reduced $\alpha$-widths
and some $\beta$-strength parameters were left free. Eventually
all parameters were varied except for the IAS reduced $\alpha$-width, $\gamma_9$, 
which in the literature has the upper limit $\Gamma_\text{tot} \le
0.3\, \text{keV}$. Due to lack of sensitivity to such a small width
$\gamma_9$ was kept at $-6.09 \times 10^{-3}\, \sqrt{\text{MeV}}$
corresponding to $\Gamma_\text{tot} = 0.200\, \text{keV}$ throughout
the minimization. To reproduce the data it was necessary to include a
broad level at 8\,767(23)~keV; we shall comment on this below.

\begin{table*}
\caption{\label{tab:results} Results expressed as observed $R$-matrix parameters $E_\lambda$, $\Gamma_\alpha$ and $B_{F/GT}$ with comparison to literature values.}
\begin{ruledtabular}
\begin{tabular}{ccccccccc} 
 &\multicolumn{3}{c}{This work}&&\multicolumn{2}{c}{Compilation \cite{tunl_A20}} && Reference \cite{clifford89} \\ \cline{2-4} \cline{6-7} \cline{9-9} 
 Level&E[MeV]&$\Gamma_\alpha$[keV]&$\text{B}_{F/GT}$
&&E[MeV] & $\Gamma_\text{total}$[keV]&&$\text{B}_{F/GT}$ \vspace{0.05cm}  \\ \hline
$5$ &$7.4227(15)$&$10.0(5)$ &$0.246(4)$&&$7.4219(12)$ & $15.1(7)$ && $0.240(8)$ \\
$6$ & $7.8301(14)$ & $0.18(8)$\footnote{The systematic uncertainty is larger than the quoted fit uncertainty.} & $0.017(7)$&& $7.8334(15)$ & $2$  && $0.0126(5)$ \\
$7$ & $8.767(23)$ & $686(56)$ & $0.00230(7)$ && $9.00(18)$ & $800$ && $0.008(12)$\footnote{In \cite{clifford89} the level is located at 9.196(30)~MeV.} \\
$8$ & $9.4927(19)$ & $35(3)$ & $0.0329(8)$ &&  $9.483(3)$ & $29(15)$ && $0.032(2)$ \\
$9$ & $10.274(2)$ & $0.2$ & $1.95(3)$ && $10.2732(19)$ & $\leq 0.300$ && $2.06(6)$ \\
$10$ & $10.587(2)$ & $34(2)$ & $0.069(2)$ && $10.584(5)$ & $24$ && $0.067(5)$ \\
$11$ & $10.842(2)$ & $16.4(5)$ & $0.258(5)$ && $10.843(4)$ & $13$ && $0.224(15)$ \\
$12$ & $11.331(3)$ & $85(6)$ & $0.127(4)$ && $11.320(9)$ & $40(10)$ && $0.101(14)$ \\
$13$ & $11.89(5)$ & $252(70)$ & $0.11(10)$ && $11.885(7)$ & $46$ && $0.039(19)$ \\
 \end{tabular}
\end{ruledtabular}
\end{table*}

The uncertainties on the $R$-matrix parameters
$\sigma\left(E_\lambda\right)$, $\sigma\left(\gamma_\lambda\right)$
and $\sigma\left(g_{\lambda,\text{F/GT}}\right)$ were calculated
by adding their individual errors in quadrature. The error originating
from the response function was estimated by comparing fits using
two different response functions: One that includes the low energy
tail of the IAS and one that does not.
The detailed error budget is as follows.

\textbf{Energy parameters ($E_{\lambda}$):} Errors on $E_{\lambda}$ from the energy calibration range
from $1.4\, \text{keV}$ to $3.0\, \text{keV}$ while the statistical
fit errors range from $2\times 10^{-3}\, \text{keV}$ to $4.5\,
\text{keV}$. The response function errors on $E_\lambda$ are in
general small but relatively large for levels 7 and 13.

\textbf{Reduced width parameters ($\gamma_{\lambda}$):} The response errors dominate
since the reduced widths are highly sensitive to the detector
response. The errors are around $2\,$-$\,4 \times 10^{-3}\,
\sqrt{\text{keV}}$ for all levels except level 7 for which it is
$2\times 10^{-2}\, \sqrt{\text{keV}}$. The statistical errors are at
the $10^{-3}\, \sqrt{\text{keV}}$ level. The uncertainty on
$\Gamma_{\text{tot},\lambda}$ is found directly by combining
$\sigma(\gamma_\lambda)$ and eq. (\ref{eq:Gamma}).

$\boldsymbol{\beta}$\textbf{-strength parameters (g$_{\lambda,\text{G/FT}}$):} The statistical
errors dominate giving contributions around $10^{-3}$ for all levels
apart from levels 6, 12 and 13 where they are one order of magnitude
larger. The uncertainty on $\text{B}_\text{F/GT}$ is obtained through
propagation of errors in eq. (\ref{eq:BGT}) where
$\sigma\left(\text{BR}_\alpha\right)$ is the dominating source of
uncertainty.

\subsubsection{Discussion}
Our observed level parameters in Table \ref{tab:results} will now be
discussed and compared to the most recent $^{20}$Ne evaluation
\cite{tunl_A20}. Compared to earlier work we propose a new
interpretation of the $\beta$-delayed $\alpha$-spectrum from
$^{20}$Na.

The energy positions of levels 5, 9, 10, 11, 12 and 13 found in this
work are in agreement with the literature values within one standard
deviation. Levels 6, 7 and 8 differ slightly more, we note that their
energy region contains clear interference effects.

The literature widths in Table \ref{tab:results} are total widths,
many are quite uncertain so that a quantitative comparison to our
values is not possible. For levels 5, 7, 8, 10, 11 and 12 our widths
are more precise than the literature ones. The width of level 5
differs significantly from the one in \cite{tunl_A20} but agrees well
with a newer value of $9.0(13)\, \text{keV}$ \cite{Costantini}. In the
case of level 6 the value of $180(80)\, \text{eV}$ is too small to be
trusted and our poor determination for
level 13 is attributed to the low level of statistics in this part of
the experimental spectrum.

The calculated $B_{\text{F/GT}}$ values are compared to those from
Clifford \textit{et al.}, which were obtained by conversion from the
reported log\,$ft$ values. Agreement within one standard deviation is
only obtained for levels 5, 6, 8 and 10 albeit level 9 nearly
agrees. The large difference for level 13 is again ascribed to the few
events collected in that region while the disagreement for levels 7, 9,
11 and 12 largely has to do with our reinterpretation of the
$\alpha$-spectrum.

The analysis performed by Clifford \textit{et al.} does not take
interference effects into account. The events between level 11 and 12
in their interpretation therefore suggest the presence of two new
levels in $^{20}$Ne (level 10 and 11 in figure 4 in
\cite{clifford89}), levels that have not been seen in other
reactions. Our analysis, which treats interference effects properly,
shows that constructive interference between levels 11 and 12
naturally reproduces the spectral shape in this region.

\begin{figure}[h]
\includegraphics[width=0.48\textwidth]{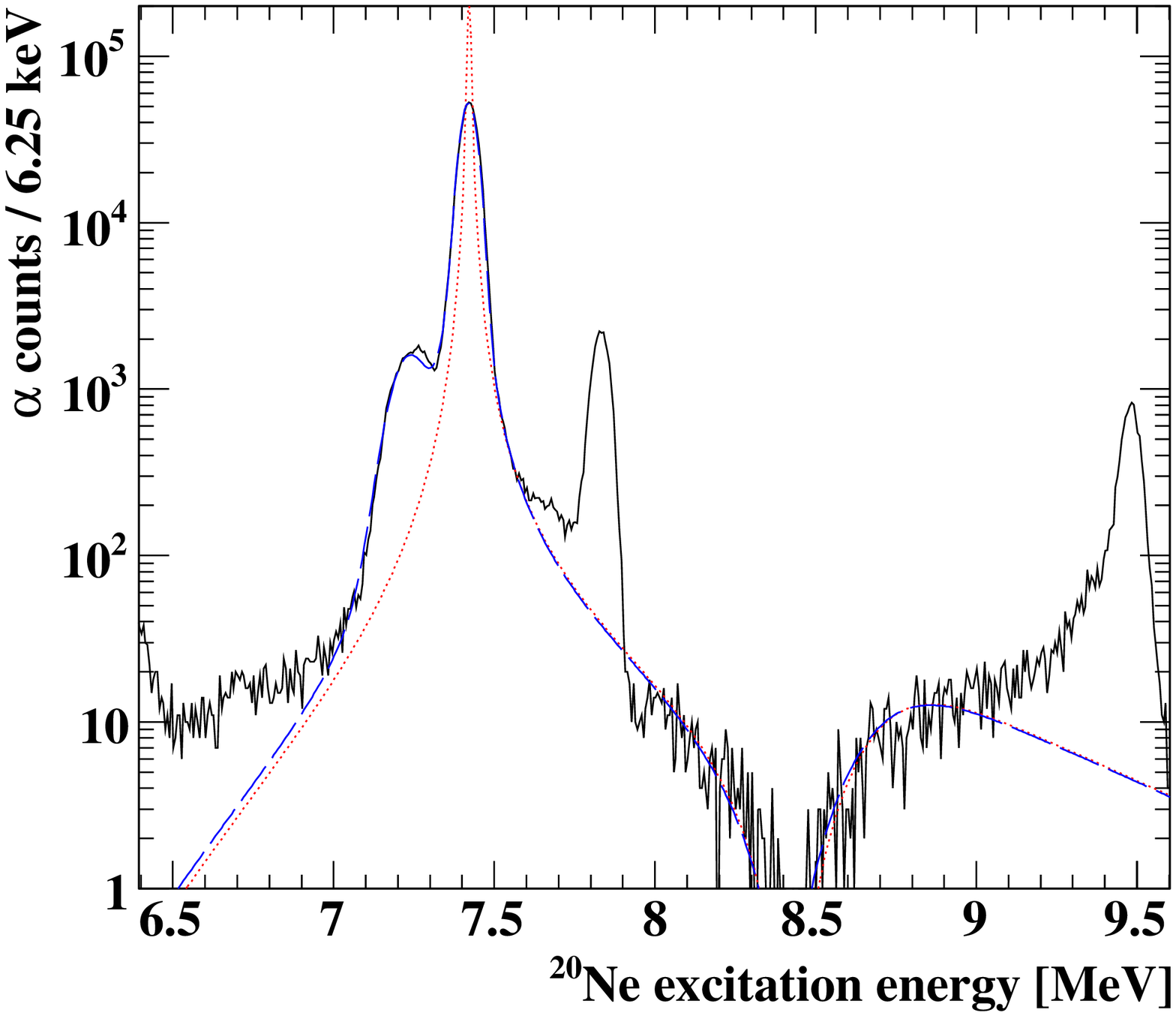}
\caption{\label{fig:fit_2level}The $R$-matrix expression
  (eq. \ref{eq:N(E)}) including only levels 5 and 7 plottet against
  data. This is \textbf{not} a fit, but the best-fit parameters for the
  two levels are used. Shown is the R-matrix expression (eq. \ref{eq:N(E)}) (dashed),
   the convoluted expression (long-dashed) and the data (solid).}
\end{figure}

Clifford \textit{et al.} explained the characteristic structure around 8\,--\,9~MeV by postulating the existence of two new levels in
$^{20}$Ne at $8\,058\, \text{keV}$ and $9\,196\, \text{keV}$, but they
do mention that interference between the $9\,196\, \text{keV}$ level
and other levels could explain the 8--9~MeV structure. Our analysis
does not include the $8\,058\, \text{keV}$ level (only reported by
Clifford \textit{et al.}).  The interference effects are here very
important and are mainly due to destructive interference between level
5 and 7. This is shown explicitly in figure \ref{fig:fit_2level} where
eq. (\ref{eq:N(E)}) including only levels 5 and 7 is plotted.

Clifford \textit{et al.} briefly mention the possibility that the 9\,196~keV level identified by them could be identical to a known broad level at 8\,800~keV. Indeed a level with $2^+$ had been observed at
$\approx8\,800\,\text{keV}$ both in $^{16}\text{O}(\alpha,\alpha)$
\cite{McDermott} and $^{19}\text{F}\left(^{3}\text{He,d}\right)$
\cite{Fortune1976} reaction experiments. Later Burlein \textit{et
al.} \cite{Burlein} measured the $^{20}\text{Ne}(\pi,\pi')$ reactions,
reported $\text{E} = 9\,000(200)\, \text{keV}$ and $\Gamma = 800\,
\text{keV}$ for this level and identified it as the $2^+$ member of
the $0_4^+$ rotational band in $^{20}$Ne. The present result,
$\text{E} = 8\,770(20)\, \text{keV}$ and $\Gamma = 686(56)\, \text{keV}$,
is the most precise observation of the $2_4^+$ level and the first
conclusive observation in $\beta$-decay.

The pronounced $^{16}\text{O}\,$+$\,\alpha$ cluster structure of
$^{20}$Ne has been verified both experimentally [35-37]
and theoretically \cite{Kimura2004} and several rotational bands containing resonances
with very large $\alpha$ widths have been identified. The properties
determined for the $8\,770\,\text{keV}$ resonance in our work clearly
indicate cluster structure. Its large $\alpha$-width points to a
substantial overlap with the exit channel wave function indicating a
large $^{16}\text{O} + \alpha$ component. More quantitatively
$\gamma_7^2 = 0.320\,\text{MeV}$ is close to the Wigner limit
$\gamma_\text{w}^2$ of $0.344\, \text{MeV}$, evaluated with a radius of
$r_0 = 1.5\left(16^{1/3} + 4^{1/3} \right)$ fm.  The cluster structure
is consistent with the small $\text{B}_\text{GT}$ value determined in
this work; an allowed Gamow-Teller transition to a pure
$^{16}\text{O}$ + $\alpha$ cluster state would be Pauli blocked.

As a final remark we note that the position for level 7 agrees fairly
well with the position found in a theoretical study \cite{Kimura2004}
for the $2_4^+$ state, thus supporting our assignment.

\subsubsection{IAS properties}
Our results give, when combined with earlier experiments, improved
information on the IAS.  First we consider the fraction of delayed
$\alpha$-particles coming from the IAS. This is best estimated from
the singles spectrum, Fig.~\ref{fig : alpha spectra}, by including the intensity above 1.5~MeV. This lower limit is varied by 100
keV to estimate the systematic uncertainty. The DSSSD detectors have a
singles $\alpha$ particle efficiency that varies slightly with energy
and has been determined in Ref.\cite{kirsebom11_8B}, the data are
corrected for this effect. Each DSSSD gives an independent value for
the ratio. The four values are consistent and lead to a final value
of $I^{\alpha}_{IAS}/I^{\alpha}_{tot} = 0.1391(5)$. Compared to the previous value of $0.1386(33)$ \cite{clifford89} the precision is improved significantly.

Our second result is the total $\beta$-$\alpha$ strength for the IAS of 1.95(3)
given in Table \ref{tab:results}. The $\beta$-$\alpha$ strength is a product of the $\beta$ strength and the branching to the $\alpha$ decay channel and must be corrected with the factor
$\Gamma^{IAS}_{\alpha}/\Gamma^{IAS}_{tot}$. This factor can be estimated in
the following way. The total Fermi strength is $2(1-\delta_c)$ where
the correction factor according to theory \cite{clifford89} is
0.014(5). Combining this with the ratio of Gamow-Teller to Fermi
strength, also measured in \cite{clifford89}, gives a value for the
total beta strength to the IAS that leads to
$\Gamma^{IAS}_{\alpha}/\Gamma^{IAS}_{tot} = 0.92(3)$. 

Alternatively, one may determine this ratio by measuring the
$\gamma$-decays of the IAS. Several branches have been seen
\cite{fifield77}, the main one goes to the 1\,630~keV level and
constitute 88.9(5)\% of all IAS $\gamma$-decays. (This number includes
only observed transitions; as yet unobserved branches, such as the one to
the 6\,725~keV level discussed above, are not expected to change the
value significantly.) Two relative intensities, that of the IAS
transition to the 1\,630~keV level relative to the total number of 1\,630
keV $\gamma$-rays \cite{ingalls76a} and that of the total number of
$\alpha$-particles relative to the 1\,630~keV $\gamma$-ray
\cite{clifford89}, now suffice to extract
$\Gamma^{IAS}_{\alpha}/\Gamma^{IAS}_{tot} = 0.961(5)$, a value more
precise than the above one. 

Finally, one may include the measurements of $\alpha$-radiative
capture through the IAS to the 1\,630~keV state
\cite{ingalls76,fifield77,Steck78}, which give an average value of $5
\Gamma^{IAS}_{\gamma\rightarrow1.63}\Gamma^{IAS}_{\alpha}/\Gamma^{IAS}_{tot}$
of 19.5(1.5) eV \cite{tunl_A20}, to put the relative measurements on an
absolute scale. This gives a $\Gamma^{IAS}_{tot}$ of 116(17) eV, which
in hindsight justifies our use above of the 116(20) eV deduced in
\cite{ingalls76}. The difference to the value of 200 eV used in the
fit (Table \ref{tab:results}) will not affect the fit quality.